\documentstyle[11pt,amsmath,amsfonts]{article}

\begin{document}

\def\Ket#1{|#1\rangle}

\newcommand{\ba}{\mbox{\boldmath $a$}}
\newcommand{\g}{\mbox{\boldmath $g$}}
\newcommand{\gs}{\mbox{\boldmath $\scriptstyle g$}}
\newcommand{\x}{\mbox{\boldmath $x$}}
\newcommand{\xs}{\mbox{\boldmath $\scriptstyle x$}}
\newcommand{\y}{\mbox{\boldmath $y$}}
\newcommand{\ys}{\mbox{\boldmath $\scriptstyle y$}}
\newcommand{\z}{\mbox{\boldmath $z$}}
\newcommand{\zs}{\mbox{\boldmath $\scriptstyle z$}}
\newcommand{\J}{\mbox{\boldmath $J$}}
\newcommand{\V}{\mbox{\boldmath $V$}}
\newcommand{\vv}{\mbox{\boldmath $v$}}
\newcommand{\Tr}{\mbox{\rm\small Tr\ }}
\newcommand{\Beta}{\mbox{B}}
\newcommand{\B}{\mbox{B}}
\newcommand{\bra}{\langle}
\newcommand{\ket}{\rangle}
\newcommand{\balpha}{\mbox{\boldmath $\alpha$}}
\newcommand{\balphas}{\mbox{\boldmath $\scriptstyle \alpha$}}
\newcommand{\bbeta}{\mbox{\boldmath $\beta$}}
\newcommand{\bbetas}{\mbox{\boldmath $\scriptstyle \beta$}}

\newcommand{\bgamma}{\mbox{\boldmath $\gamma$}}
\newcommand{\bdelta}{\mbox{\boldmath $\delta$}}
\newcommand{\bsigma}{\mbox{\boldmath $\sigma$}}
\newcommand{\bxi}{\mbox{\boldmath $\xi$}}
\newcommand{\bxis}{\mbox{\boldmath $\scriptstyle\xi$}}
\newcommand{\bnabla}{\mbox{\boldmath $\nabla$}}
\newcommand{\one}{\mbox{\tt 1}\hspace{-0.057 in}\mbox{\tt l}}
\newcommand{\Z}{\mbox{\tt Z}\hspace{-0.057 in}\mbox{\tt Z}}
\newcommand{\diverg}{\mbox{\rm div}}
\def\smiley{\hbox{\large$\bigcirc$\hspace{-.80em}%
\raise.2ex\hbox{$\cdot\cdot$}\kern-.61em   
\lower.2ex\hbox{\scriptsize$\smile$}}\ }
\def\angry{\hbox{\large$\bigcirc$\hspace{-.80em}%
\raise.2ex\hbox{$\cdot\cdot$}\kern-.61em   
\lower.2ex\hbox{\scriptsize$\sim$}}\ }
\def\frowney{\hbox{\large$\bigcirc$\hspace{-.80em}%
\raise.2ex\hbox{$\cdot\cdot$}\kern-.635em  
\lower.2ex\hbox{\scriptsize$\frown$}}\ }
\def\blahey{\hbox{\large$\bigcirc$\hspace{-.80em}%
\raise.2ex\hbox{$\cdot\cdot$}\kern-.46em    
\lower.2ex\hbox{\scriptsize\hbox{--}}}\ }

\title{Classical limit in terms of symbolic dynamics for the quantum
baker's map}

\author{Andrei N. Soklakov\footnote{~a.soklakov@rhul.ac.uk}~~and R\"udiger
 Schack\footnote{~r.schack@rhul.ac.uk}\\
\\
{\it  Department of Mathematics, Royal Holloway, University of London, }\\
{\it Egham, Surrey TW20 0EX, UK }\\
{\it and} \\
{\it Center for Advanced Studies, Department of Physics and Astronomy,} \\
{\it University of New Mexico, Albuquerque, New Mexico 87131--1156, USA.}}

\date{12 August 1999}

\maketitle

\begin{abstract}
We derive a simple closed form for the matrix elements of the quantum baker's
map that shows that the map is an approximate shift in a symbolic
representation based on discrete phase space. We use this result to give a
formal proof that the quantum baker's map approaches a classical Bernoulli
shift in the limit of a small effective Planck's constant.
\end{abstract}

\section{Introduction}

The quantum baker's map \cite{Balazs1989,Saraceno1990} is a
prototypical quantum map invented for the theoretical study of quantum
chaos. During the last decade, its semiclassical properties have been
studied extensively
\cite{Ozorio1991,Dittes1994,Saraceno1994a,Lakshminarayan1995,%
Kaplan1996,DaLuz1995}, it has been shown to display
hypersensitivity to perturbation \cite{Schack1993e,Schack1996b},
optical \cite{Hannay1994} and quantum computing
\cite{Schack1998a,Brun1999a} realizations have been proposed, its
long-time behavior has been investigated \cite{Dittes1994,Kaplan1996},
it has been studied in a path-integral approach \cite{DaLuz1995} and
defined on a sphere \cite{Pakonski1999}.  The quantum baker's map is a
quantized version of the classical baker's transformation
\cite{Arnold1968}, but there is no unique quantization procedure
\cite{Berry1979}. The original definition of the map
\cite{Balazs1989,Saraceno1990} is based on Weyl's quantization
\cite{Weyl1950} of the unit square. Essentially the same map has been
derived by algebraic methods \cite{Rubin1998a,Lesniewskietall} as well as by
considering the transition from ray to wave optics \cite{Hannay1994}. 
Recently a whole
class of quantum baker's maps has been defined \cite{AAECC} by
exploiting formal similarities between the symbolic dynamics
\cite{Alekseev1981} for the classical map on the one hand and the
dynamics of strings of quantum bits of the type considered in the theory of
quantum computing on the other hand. This class of quantum baker's
maps, which can also be derived from the semiquantum maps introduced in
Ref.~\cite{Saraceno1994a}, is the subject of this paper.

The classical baker's transformation \cite{Arnold1968} maps the unit square $0
\leq q,p \leq 1$ onto itself according to
\begin{equation}
(q,p) \longmapsto \left\{  \begin{array}{ll}
\Bigl(2q,{1\over2}p\Bigr)\;,       &\mbox{if $0\leq q\leq{1\over2}$,} \\
\Bigl(2q-1,{1\over2}(p+1)\Bigr)\;, &\mbox{if ${1\over2}<q\leq1$.}
                           \end{array}  \right. 
\label{eqcbaker}
\end{equation}
This corresponds to compressing the unit square in the $p$ direction and
stretching it in the $q$ direction, while preserving the area, then cutting it
vertically, and finally stacking the right part on top of the left part---in
analogy to the way a baker kneads dough.  
The classical baker's map, has a simple description
in terms of its symbolic dynamics \cite{Alekseev1981}. Each point
$(q,p)$ is represented by a symbolic string 
\begin{equation}
s = \cdots s_{-2} s_{-1} s_0 . s_1 s_2 \cdots \;,
\label{eq:biinfinite}
\end{equation}
where $s_k=0$ or $1$, and
\begin{equation}
q=\sum_{k=1}^{\infty}s_k 2^{-k} \;,\;\;\;\;
p=\sum_{k=0}^{\infty}s_{-k} 2^{-k-1} \;.
\label{eq:qp}
\end{equation}
The action of the baker's map on a symbolic string $s$ is then given
by the shift map (or Bernoulli shift) $U$ defined by $Us=s'$, where
$s'_k=s_{k+1}$. This means that, at each time step, the entire string
is shifted one place to the left while the dot remains fixed. Although
the relation (\ref{eq:qp}) between points $(q,p)$ and symbolic strings
is particular to the baker's transformation, the method of symbolic
dynamics is very general and can be applied to a large class of
chaotic maps \cite{Alekseev1981}.

Symbolic representations for the {\it quantum\/} baker's map have been
introduced in Refs.\ \cite{Saraceno1994a,Schack1998a,AAECC}. These
representations are all obtained by writing the quantum propagator
$\hat B$ in a {\it mixed\/} form $\langle i|\hat B|i'\rangle$, where
$\{\Ket{i}\}$ and $\{\Ket{i'}\}$ are different bases. In this paper,
we derive a simple closed form of the matrix elements with respect to
a single basis. We show that all members of the class of quantum
baker's maps defined in Ref.~\cite{AAECC} are approximate shifts in a symbolic
representation based on discrete phase space. We use this result to
give a formal proof that all members of this class of quantum baker's maps 
approach a
classical Bernoulli shift in the limit of a small effective Planck's
constant. 

The paper is organized as follows. In Sec.~\ref{sec:background}, we
give the necessary background and definitions. In
Sec.~\ref{sec:results}, we state the results of the paper and
discuss their significance. Finally, Sec.~\ref{sec:proofs} contains
the derivations and proofs.

\section{Background} \label{sec:background}

Most results of this paper are phrased in terms of finite binary
strings. It will be convenient to adopt a slightly different and more
flexible notation than the one used in Eq.~(\ref{eq:biinfinite}). 
Here, a binary string
\begin{equation}                                                 \label{baker1}
\xi_{s:f}\stackrel{\rm def}{=}\left\{ \begin{array}{ll} 
          \xi_s\xi_{s+1}\cdots \xi_f&\mbox{($s\leq f$)}\cr
          \xi_s\xi_{s-1}\cdots \xi_f&\mbox{($s>f$)} \;,
        \end{array}
\right.
\end{equation}
where $\xi_i\in\{0,1\}$ is a bit, can have increasing ($s<f$) or
decreasing ($s>f$) indices. We will use
bold Greek and Latin letters to denote binary strings, e.g.,
\begin{equation}                                                 \label{baker2}
\balpha=\xi_{s:f}\mbox{\ \ or\ \ }\x=x_{h:t} \;.
\end{equation}
The length of a string $\balpha$ will be denoted by $|\balpha|$; e.g.,
in the above example, $|\balpha|=|f-s|+1$.
Concatenation of strings is defined in the usual way.
Again considering the above example, $\balpha\x$ is the string
$\balpha\x=\xi_s...\xi_f x_h... x_t$.
Any string $\balpha$ represents a natural number through its binary
expansion
\begin{equation}                                                 \label{baker6}
\balpha=\sum_{k=1}^{|\balphas|}2^{|\balphas|-k}\balpha_{(k)}\;,
\end{equation}
where $\balpha_{(k)}$ denotes the $k$-th
bit of $\balpha$, $1\leq k\leq |\balpha|$, such that
\begin{equation}                                                 \label{baker5}
\balpha=\balpha_{(1)}\balpha_{(2)}...\balpha_{(|\balphas|)} \;.
\end{equation}
Thus our notation does not distinguish between a binary string and the
corresponding natural number. Similarly, two strings $\balpha$ and $\x$
can be combined to represent a rational number
\begin{equation}                                                 \label{baker7}
\balpha.\x\stackrel{\rm def}{=}
              \sum_{k=1}^{|\balphas|}2^{|\balphas|-k}\balpha_{(k)}
                                +\sum_{k=1}^{|\xs|}2^{-k}\x_{(k)} \;.
\end{equation}

Quantum baker's maps are defined on the $D$-dimensional Hilbert space
of the quantized unit square \cite{Weyl1950}. For consistency of units,
we let the quantum scale on ``phase space'' be $2\pi\hbar=1/D$.
Following Ref.~\cite{Saraceno1990}, we choose half-integer eigenvalues
$q_j=(j+{1\over2})/D$, $j=0,\ldots,D-1$, and $p_k=(k+{1\over2})/D$,
$k=0,\ldots,D-1$, of the discrete ``position'' and ``momentum''
operators $\hat q$ and $\hat p$, respectively, corresponding to
antiperiodic boundary conditions.  We further assume that $D=2^N$,
which is the dimension of the Hilbert space of $N$ qubits, i.e., $N$
two-state systems.

The $D=2^N$ dimensional Hilbert space modeling the unit square can be 
realized as the product space of $N$ qubits in such a way that
\begin{equation}
\Ket{q_j} =
\Ket{\xi_1}\otimes\Ket{\xi_2}\otimes\cdots\otimes\Ket{\xi_N}  \;,
\label{eq:tensor1}
\end{equation}
where $j=\sum_{l=1}^N \xi_l2^{N-l}$, $\xi_l\in\{0,1\}$,
and where each qubit has basis states $|0\rangle$ and $|1\rangle$.
It follows that, written in our string notation as binary numbers, 
$j=\xi_1\xi_2\ldots \xi_N=\xi_{1:N}$ and 
$q_j=0.\xi_1\xi_2\ldots \xi_N1=0.\xi_{1:N}1$. We define the notation 
\begin{equation}
\Ket{.\xi_{1:N}}=\Ket{.\xi_1\xi_2\ldots \xi_N} = e^{i\pi/2} \Ket{q_j} \;,
\label{eqtensor}
\end{equation}
which is closely analogous to Eq.~(\ref{eq:biinfinite}), where the
bits to the right of the dot specify the position variable; see
Ref.~\cite{AAECC} for the reason for the phase shift $e^{i\pi/2}$.

Momentum and position eigenstates are related through the quantum Fourier
transform operator $\hat F$ {\cite{Saraceno1990}}, i.e., 
$\hat F\Ket{q_k}=\Ket{p_k}$.
Again in analogy to Eq.~(\ref{eq:biinfinite}), we define the notation
$|\xi_{1:N}.\ket=\Ket{p_k}$, where $p_k=0.\xi_{N:1}1$.

By applying a {\it partial\/}
quantum Fourier transform \cite{AAECC} to the position eigenstates,
one obtains the family of states
\begin{eqnarray}                                                 \label{baker8}
|\xi_{1:n}.\xi_{n+1:N}\ket&\stackrel{\rm def}{=}&|\xi_{n+1}\ket
            \otimes\cdots\otimes
            |\xi_N\ket e^{i\pi(0.\xi_{n:1}1)}\otimes\cr
                  & & \sqrt{1/2}\{|0\ket+\exp[2\pi i(0.\xi_11)]|1\ket\}
                       \otimes\cr
                  & & \sqrt{1/2}\{|0\ket+\exp[2\pi i(0.\xi_2\xi_11)]|1\ket\}
                      \otimes\cdots\otimes\cr
                  & & \sqrt{1/2}\{|0\ket+\exp[2\pi i(0.\xi_{n:1}1)]|1\ket\}\;,
\end{eqnarray}
where $1\leq n\leq N-1$.
More precisely, the state $|\xi_{1:n}.\xi_{n+1:N}\ket$ is obtained by 
applying the Fourier transform operator to the $n$ rightmost bits of
the position eigenstate $|.\xi_{n+1:N}\xi_{n:1}\ket$. 
For given $n$, these states form an orthogonal basis. 
The state $|\xi_{1:n}.\xi_{n+1:N}\ket$ is localized in both 
position and momentum: it is strictly localized within a position region 
of width $1/2^{N-n}$, centered at position 
$q=0.\xi_{n+1:N}1$, and it is crudely localized within 
a momentum region of width $1/2^{n}$, centered at momentum 
$p=0.\xi_{n:1}1$.

For each $n$, $1\leq n\leq N-2$, a quantum baker's map can be defined by
\begin{equation}                                                 \label{baker9}
\hat{B}|\xi_{1:n}.\xi_{n+1:N}\ket\stackrel{\rm def}{=}
                    |\xi_{1:n+1}.\xi_{n+2:N}\ket \;,
\end{equation}
where the dot is shifted by one position.  In phase-space language,
the map $\hat B$ takes a state localized at
$(q,p)=(0.\xi_{n+1:N}1,0.\xi_{n:1}1)$ to a state localized at
$(q',p')=(0.\xi_{n+2:N}1,0.\xi_{n+1:1}1)$, while it stretches the
state by a factor of two in the $q$ direction and squeezes it by a
factor of two in the $p$ direction. This analogy with the classical
baker's map motivates calling the maps (\ref{baker9}) ``quantum
baker's maps.''  For $n=N-1$, the map is the original quantum baker's map
as defined in Ref.~\cite{Saraceno1990}, which in our notation becomes
\begin{equation}                      \label{eq:mapbig}
\hat{B}|\xi_{1:N-1}.\xi_N\ket\stackrel{\rm def}{=}
                    |\xi_{1:N}.\ket \;,
\end{equation}
and for $n=0$, the map is
\begin{equation}                      \label{eq:mapsmall}
\hat{B}|.\xi_{1:N}\ket\stackrel{\rm def}{=}
                    |\xi_{1}.\xi_{2:N}\ket \;.
\end{equation}
Below we show that all the maps
(\ref{baker9},\ref{eq:mapbig},\ref{eq:mapsmall}) 
reduce to the classical baker's map in the limit $\hbar\rightarrow0$.

\section{Results} \label{sec:results}

Equation (\ref{baker9}) is a mixed representation of the quantum
baker's map, using different bases on both sides of the equation.
To go beyond the heuristic phase-space interpretation of the map given
at the end of the last section, we need to express the matrix elements
of $\hat B$ with respect to a single basis, i.e., we need to find
\begin{equation}                                                \label{baker10}
C^{\rm 1st}(\bxi^0,\bxi^1) \stackrel{\rm def}{=} \left\{ \begin{array}{ll}
\langle.\xi_{1:N}^1| \hat B |.\xi_{1:N}^{0}\ket
& \mbox{if $n=0$} \\
\langle\xi_{1:n}^1.\xi_{n+1:N}^1| \hat B |\xi_{1:n}^{0}.\xi_{n+1:N}^{0}\ket
& \mbox{if $1\leq n\leq N-1$} \;,
\end{array} \right.
\end{equation}
where $\bxi^0{=}\xi_{1:N}^{0}$ and
$\bxi^1{=}\xi_{1:N}^1$. A main result of this paper
is the following simple formula, which will be proved in Sec.~\ref{sec:proofs}:
\begin{equation}                                                \label{baker11}
C^{\rm 1st}(\bxi^0,\bxi^1)
=\Phi(\xi^0_1,\xi_N^1)
      \frac{\delta(\xi^0_{n+2:N}-\xi^1_{n+1:N-1})}{2^{n+1}\sin
                  [\pi(0.\xi_{n+1:1}^01-0.\xi_{n:1}^11)]}\;,
\end{equation}
where $1\leq n\leq N-2$ and $\Phi$ is a phase factor given by  
\begin{equation}                                                \label{baker12}
\Phi(\xi^0_1,\xi^1_N)=\frac{1}{\sqrt{2}}
                      [i(-1)^{\xi_N^1}-(-1)^{\xi_1^0}]\;.
\end{equation}
For the case $n=0$, one obtains
\begin{equation}                                          \label{eq:nsmall}
C^{\rm 1st}(\bxi^0,\bxi^1)
=\frac{1-i}{2}\delta(\xi^0_{2:N}-\xi^1_{1:N-1})
               e^{i\frac{\pi}{2}|\xi^0_1-\xi^1_N|} \;,
\end{equation}
and for $n=N-1$,
\begin{equation}                                  \label{eq:nbig}
C^{\rm 1st}(\bxi^0,\bxi^1)
= \frac{\Phi(\xi^0_1,\xi_N^1)}{2^{N}\sin
                  [\pi(0.\xi_{N:1}^01-0.\xi_{N-1:1}^11)]}\;.
\end{equation}
The coefficients $C^{\rm 1st}(\bxi^0,\bxi^1)$ given in (\ref{baker11})
are zero unless 
$\xi^0_{n+2:N}=\xi^1_{n+1:N-1}$, i.e., unless the position bits 
$\xi^1_{n+1:N-1}$ of the final state are obtained by shifting the
corresponding position bits of the initial state.
Furthermore, the $\sin$ term in the denominator 
ensures that  $C^{\rm 1st}$ is strongly peaked
for $\xi^0_{n+1:2}=\xi^1_{n:1}$, i.e., if the momentum bits
$\xi^1_{n:1}$ of the final state  are obtained by shifting the
corresponding momentum bits of the initial state.
The formula (\ref{baker11}) therefore establishes rigorously that
the maps (\ref{baker9}) are approximate shift maps, a result which had
been obtained numerically in Ref.~\cite{Saraceno1994a}.

To formulate the question of the classical limit of the baker's map,
we use the concept of coarse-graining in the spirit of the
consistent (or decoherent) histories approach
\cite{Griffiths,Omnes,GellMannHartle}. For this, we introduce
projectors on subspaces corresponding to symbolic strings $\y$ of length
$l$.  We fix in advance an upper limit,
$k_{\rm max}$, on the number of iterations, $k$, considered; this is
necessary because in computing the classical limit of a chaotic map,
the limit $\hbar=2^{-(N+1)}/\pi\to0$ has to be taken before the limit $k\to\infty$ 
\cite{Berry1991}. We will show that, for given $l$ and $k_{\rm max}$,
it is always possible to choose $\hbar$ in  such a way that the coarse-grained
quantum dynamics is arbitrarily close to a shift of the string $\y$.
In contrast to the approach of
Refs.~\cite{Rubin1998a,Lesniewskietall}, in taking the limit
$\hbar\to0$, we always remain in the finite-dimensional Hilbert space
on which our maps are defined.

As before, we are considering basis states of the form
$\Ket{\xi_{1:n}.\xi_{n+1:N}}$. As we let $N$ increase, the number of
position bits to the right of the dot, $m\stackrel{\rm def}{=}N-n$,
remains fixed. We define $r=N-l$ as the number of bits ignored in the
coarse graining. In the following, we always assume that $k\leq k_{\rm max}<r$.

We are now in a position to introduce a family of projectors
\begin{equation}                    \label{baker13}
P_{\ys}^{r,k}\stackrel{\rm def}{=} \left\{ \begin{array}{ll}
\displaystyle \sum_{|\xs|=r-k,\,|\gs|=k\atop\,}|\x\y^1.\y^2\g\ket
\bra\x\y^1.\y^2\g| & \mbox{if $k< m$} \cr
\displaystyle \sum_{|\xs|=r-k,\,|\gs^2|=m\atop\,|\gs^1|=k-m}
                 |\x\y\g^1.\g^2\ket\bra\x\y\g^1.\g^2| & \mbox{if $k\geq m$}\;,
  \end{array} \right.
\end{equation}
where $\y^1\y^2=\y$ and $|\y^2|=m-k$.  By normalizing these
projectors, we obtain a family of uniform density matrices,
\begin{equation}                                                \label{baker15}
\rho_k\stackrel{\rm def}{=}2^{-r}P^{r,k}_{\ys} \;.
\end{equation}
A classical shift acts on these states as 
\begin{equation}                    
\rho_k\mapsto\rho_{k+1} \;.
\end{equation}
Projecting a state $\rho_{k'}$ onto the shifted subspace
$P_{\ys}^{r,k}$ gives the characteristic delta distribution
\begin{equation}                                                \label{baker17}
\Tr[P^{r,k}_{\ys}\rho_{k'}]=\delta_{kk'}\;.
\end{equation}
We will prove that
\begin{equation}                                                \label{baker16}
\Tr[P^{r,k}_{\ys}\hat{B}^k\rho_0(\hat{B}^{\dag})^k]=1
-O(\frac{r}{2^{r-k}})
\end{equation}
or, since $k$ is bounded from above by $k_{\rm max}$, and $r=N-l$, where
$l$ is fixed, 
\begin{equation}                                                \label{baker16a}
\Tr[P^{r,k}_{\ys}\hat{B}^k\rho_0(\hat{B}^{\dag})^k]=1
-O(\frac{N}{2^{N}})= 1-O(\hbar\log\hbar) \;.
\end{equation}
Comparing Eqs.~(\ref{baker17}) and (\ref{baker16a}), one sees  
that the coarse-grained quantum evolution approaches the
shift-map behavior to any required accuracy as $\hbar\to0$.
A measurement of the projectors $P^{r,k}_{\ys}$ can be interpreted as
a measurement in which the $r-k$ leftmost bits and the $k$ rightmost
bits of the symbolic string are not resolved.

Equation (\ref{baker16}) can be rewritten as
\begin{equation} \label{baker19}
2^{-r}\sum_{|\xs|=r}\Tr[P^{r,k}_{\ys}\hat{B}^k
|\x\y^1.\y^2\ket\bra\x\y^1.\y^2|(\hat{B}^{\dag})^k]= 1
-O(\frac{r}{2^{r-k}}) \;,
\end{equation}
which is a sum of $2^r$ terms bounded from above
as
\begin{equation}  \label{baker20}
\Tr[P^{r,k}_{\ys}\hat{B}^k
|\x\y^1.\y^2\ket\bra\x\y^1.\y^2|(\hat{B}^{\dag})^k]\leq 1\;.
\end{equation}
Here, $\y^1\y^2=\y$ and $|\y^2|=m$ as before.
Equations (\ref{baker19}) and (\ref{baker20}) can be both satisfied only
if the condition
\begin{equation} \label{baker18}
\Tr[P^{r,k}_{\ys}\hat{B}^k
|\x\y^1.\y^2\ket\bra\x\y^1.\y^2|(\hat{B}^{\dag})^k]=1
-O(\frac{r}{2^{r-k}})
\end{equation}
holds for all $\x$ except for a fraction of order $r/2^{r-k}$, i.e.,
for all basis states $\Ket{\xi_{1:n}.\xi_{n+1:N}}$ except for an
exponentially small fraction. 
In other words, the property (\ref{baker18}) holds for {\it typical\/} basis
states.

An interesting feature of the quantum baker's map is that there are
atypical basis states for which Eq.~(\ref{baker18}) does not hold.
In section IV we give an example of an atypical state
$|\x^{\rm atyp}\y^1.\y^2\ket$ for which
\begin{equation} \label{baker21}
\Tr[P^{r,1}_{\ys}\hat{B}|\x^{\rm atyp}\y^1.\y^2 \ket
\bra\x^{\rm atyp}\y^1.\y^2 |\hat{B}^{\dag}]=\frac{\pi^2+8G}{2\pi^2}
                                  +O(4^{r-n})+O(2^{-r}) \;,
\end{equation}
where $G\simeq0.915965$ is Catalan's constant \cite{GradshteynRyzhik}.  For
sufficiently large $n-r$ and $r$, this expression is less  
than $0.872$.  This is an
example where the quantum evolution in the limit $\hbar\to0$ differs
substantially from the classical evolution, already after the first iteration of
the map.  If, however, the initial state is a mixture in which atypical states
have an exponentially small weight, such as $\rho_0$, the correspondence
principle is obeyed.

\section{Derivations and proofs}  \label{sec:proofs}

\subsection{First Iteration}

In this section we prove the formula (\ref{baker11}) for the
matrix elements $C^{\rm 1st}(\bxi^0,\bxi^1)$. Equation (\ref{eq:nbig})
for the case $n=N-1$ follows from almost identical arguments, and 
Eq.~(\ref{eq:nsmall}) for $n=0$ is essentially trivial.
A direct calculation yields
\begin{eqnarray}                                                \label{baker22}
 C^{\rm 1st}(\bxi^0,\bxi^1)&=&\delta(\xi_{n+2}^0-\xi_{n+1}^1)
 \delta(\xi_{n+3}^0
                    -\xi_{n+2}^1)\cdots\delta(\xi_{N}^0-\xi_{N-1}^1)\cr
   & &\times \sqrt{1/2}\{\delta(\xi_N^0)+\delta(\xi_N^0-1)
             \exp[2\pi i(0.\xi_1^01)]\}\cr
   & &\times \exp\{i\pi(0.\xi_{n+1:1}^01-0.\xi_{n:1}^11)\} \cr
   & &\times 1/2[1+\exp\{2\pi i(0.\xi_2^0\xi_1^01-0.\xi_1^11)\}]\times\cdots\cr
   & &\times 1/2[1+\exp\{2\pi i(0.\xi_{n+1:1}^01-0.\xi_{n:1}^11)\}]\;.
\end{eqnarray}
Using the identity $1+e^{i\phi}=2e^{i\phi/2}\cos(\phi/2)$ and noticing
that $\delta(\xi_N^1)+\delta(\xi_N^1-1)\exp[2\pi i(0.\xi_1^01)]$ \\
=$\exp[i\pi\xi_N^1(\xi_1^0+1/2)]$, we have
\begin{eqnarray}                                                \label{baker23}
C^{\rm 1st}(\bxi^0,\bxi^1) &=&\sqrt{1/2}
                              \exp[i\pi\xi_N^1(\xi_1^0+1/2)]
                              \delta(\xi_{n+2:N}^0-\xi_{n+1:N-1}^1)\cr
  & &\times \exp[i\pi(0.\xi_{n+1:1}^01-0.\xi_{n:1}^11)]\cr
  & &\times \prod_{k=2}^{n+1}\cos[\pi(0.\xi^0_{k:1}1-0.\xi_{k-1:1}^11)]\cr
  & &\times \prod_{k=2}^{n+1}\exp[i\pi(0.\xi^0_{k:1}1-0.\xi_{k-1:1}^11)]\cr
  &=&\sqrt{1/2}\exp[i\pi\xi_N^1(\xi^0_1+1/2)]\delta(\xi^0_{n+2:N}-
                                                        \xi_{n+1:N-1}^1)\cr
  & &\times e^{i\phi_n}(\prod_{k=1}^{n}\cos\phi_k)(\prod_{k=1}^{n}e^{i\phi_k})\;,
\end{eqnarray}
where
\begin{equation}                                                \label{baker24}
\phi_k\stackrel{\rm def}{=}\pi(0.\xi^0_{k+1:1}1-0.\xi^1_{k:1}1)\;.
\end{equation}

To simplify Eq.~(\ref{baker23}), we first consider
the products of cosines and exponents separately and then combine them to
formulate the final result for the first iteration of the baker's map.
Note that
\begin{eqnarray}                                                \label{baker25}
 2\phi_k &=& \pi(0.\xi^0_{k:1}1-0.\xi^1_{k-1:1}1)+
             \pi(\xi^0_{k+1}-\xi_k^1) \cr
         &=& \phi_{k-1}+\pi(\xi^0_{k+1}-\xi_k^1)\;,
\end{eqnarray}
so
\begin{eqnarray}                                                \label{baker26}
 \cos\phi_{k-1}&=&\cos[2\phi_k+\pi(\xi_k^1-\xi^0_{k+1})]\cr
                 &=&(-1)^{\xi_k^1-\xi^0_{k+1}}\cos(2\phi_k) \;,\;\;\;k\leq n\;.
\end{eqnarray}
{From} Eq.~(\ref{baker25}), we have $2\phi_k= 4\phi_{k+1}$ (mod $2\pi$) and thus
$2\phi_k= 2^{n+1-k}\phi_n$ (mod $2\pi$), so the previous formula can be
rewritten as
\begin{eqnarray}                                                \label{baker27}
\cos\phi_k&=&(-1)^{\xi_{k+1}^1-\xi_{k+2}^0}\cos(2^{n-k}\phi_n)
\;,\:\:\:k\leq n-1\;.
\end{eqnarray}
Using this formula the product of cosines can be expressed as
\begin{eqnarray}                                                \label{baker28}
\prod_{k=1}^{n}\cos\phi_k&=&\cos\phi_n\prod_{k=1}^{n-1}\cos\phi_k \cr
    &=&(-1)^{\sigma(\xi^1_{2:n})-\sigma(\xi^0_{3:n+1})}\prod_{k=1}^{n}\cos[2^{-k}(2^n\phi_n)]\;,
\end{eqnarray}
where $\sigma(\xi_{k:n})\stackrel{\rm def}{=}\sum_{s=k}^n \xi_s$.
It is easy to check by induction that
\begin{equation}                                                \label{baker29}
\prod_{k=0}^{n-1}\cos 2^kx=\frac{\sin 2^nx}{2^n\sin x}\;,
\ \ \ x\neq \pi j\;,\ \ \ j=0,\pm 1,\pm 2, \ldots
\end{equation}
In our case
\begin{equation}                                                \label{baker30}
     \prod_{k=1}^{n}\cos[2^{-k}(2^n\phi_n)]
      =\frac{\sin(2^n\phi_n)}{2^n\sin\phi_n}\;.
\end{equation}
Putting everything together, the product of cosines becomes
\begin{equation}                                                \label{baker31}
\prod_{k=1}^{n}\cos\phi_k=(-1)^{\sigma(\xi^1_{2:n})-\sigma(\xi^0_{3:n+1})}\,
                          \frac{\sin(2^n\phi_n)}{2^n\sin\phi_n}\;,
\end{equation}
where $\phi_n=\pi(0.\xi^0_{n+1:1}1-0.\xi_{n:1}^11)$.
Now we simplify the product of exponents in (\ref{baker23}).
Equation~(\ref{baker25}) implies
\begin{equation}                                                \label{baker32}
\phi_{n-k}=2^k\phi_n+\sum_{s=1}^{k}2^{k-s}
           \pi(\xi^1_{n+1-s}-\xi^0_{n+2-s}),\mbox{\ }k\geq 1\;,
\end{equation} 
so 
\begin{eqnarray}                                                \label{baker33}
\sum_{k=1}^n\phi_k&=&\phi_n+\sum_{k=1}^{n-1}\phi_{n-k}\cr
                  &=&\phi_n\sum_{k=0}^{n-1}2^k +\sum_{k=1}^{n-1}\sum_{s=1}^k
                      2^{k-s}\pi(\xi^1_{n+1-s}-\xi^0_{n+2-s})\cr
                  &=&[\phi_n(2^n-1)+\sum_{k=1}^{n-1}\pi(\xi^1_{n+1-k}-
                           \xi^0_{n+2-k})](\mbox{mod $2\pi$})\cr
                  &=&\{\phi_n(2^n-1)+\pi[\sigma(\xi^1_{2:n})-\sigma(\xi^0_{3:n+1})]\}
                     (\mbox{mod $2\pi$})\;.
\end{eqnarray}
The product of exponents is thus given by
\begin{equation}                                                \label{baker34}
\prod_{k=1}^{n}e^{i\phi_k}=\exp(i\sum_{k=1}^{n}\phi_k)=
                           (-1)^{\sigma(\xi^1_{2:n})
                           -\sigma(\xi^0_{3:n+1})}
                           \frac{\exp(i2^n\phi_n)}{\exp(i\phi_n)}\;.
\end{equation}
Using (\ref{baker31}) and (\ref{baker34})
one can rewrite (\ref{baker23}) as
\begin{equation}                                                \label{baker35}
C^{\rm 1st}(\bxi^0,\bxi^1)
            =\frac{(-1)^{\xi_N^1(\xi^0_1+1/2)}}{2^{n+1/2}}
             \delta(\xi^0_{n+2:N}-\xi^1_{n+1:N-1})
             \frac{\sin(2^n\phi_n)}{\sin\phi_n}
             \exp(i2^n\phi_n)\;.
\end{equation}
Further simplification is possible due to the fact that $2^n\phi_n=
2\pi(0.\xi^0_2\xi^0_11-0.\xi^1_11)$ (mod $2\pi$). The final result is
\begin{equation}                                                \label{baker36}
C^{\rm 1st}(\bxi^0,\bxi^1)
=\Phi(\xi^0_1,\xi_N^1)
      \frac{\delta(\xi^0_{n+2:N}-\xi^1_{n+1:N-1})}{2^{n+1}
                   \sin[\pi(0.\xi^0_{n+1:1}1-0.\xi_{n:1}^11)]}\;,
\end{equation}
where the phase factor $\Phi$ is  given by
\begin{eqnarray}                                                \label{baker37}
\Phi(\xi^0_1,\xi_N^1)&=&
\sqrt{2}e^{i\pi\xi_N^1(\xi_1^0+1/2)}
\sin(2^n\phi_n)\exp(i2^n\phi_n)\cr
&=&\frac{1}{\sqrt{2}}[i(-1)^{\xi^1_N}-(-1)^{\xi^0_1}]\;.
\end{eqnarray}
This formula is an exact expression for the matrix elements
 (\ref{baker10})  of the quantum baker's map.

\subsection{$k$-th iteration}

In this section we prove that for all $k\leq k_{\rm max}$,
\begin{equation}                                                \label{baker38}
\Tr[P^{r,k}_{\ys}\hat{B}^k\rho_0(\hat{B}^{\dag})^k]=1
-O(\frac{r}{2^{r-k}})\;,
\end{equation}
where the projectors $P^{r,k}_{\ys}$ and the density operators
$\rho_j$  are defined in Eqs.~(\ref{baker13}) and (\ref{baker15}).
The first step is to prove that
\begin{equation}                                                \label{baker39}
\Tr[P^{r,k}_{\ys}\hat{B}\rho_{k-1}\hat{B}^{\dag}]=1-O(\frac{r}{2^{r-k}})\;.
\end{equation}
By a direct calculation, we obtain
\begin{eqnarray}                                                \label{baker40}
\Tr[P^{r,k}_{\ys}\hat{B}\rho_{k-1}\hat{B}^{\dag}]&=&
2^{k-r}\sum_{|\balphas|=r-k+1\atop{|\bbetas|=r-k\atop\,}}
|2^{n+1}\sin[2^{-(n-r+k)}\pi(0.\balpha1-0.\bbeta1)]|^{-2}\cr
&\geq& \frac{4}{\pi^22^{r-k}}\sum_{u=0}^{2^{r-k+1}-1}\sum_{v=0}^{2^{r-k}-1}
(2u-4v-1)^{-2}\;,
\end{eqnarray}
where the inequality $\sin^2x\leq x^2$ was used.
Let $L=r-k$, and let $Q(s)$ be the number of different
pairs $(u,v)$, $0\leq u<2^{L+1}$, $0\leq v<2^{L}$, for which 
$u-2v=s$. It follows that
\begin{eqnarray}                                                \label{baker41}
\sum_{u=0}^{2^{r-k+1}-1}\sum_{v=0}^{2^{r-k}-1}
(2u-4v-1)^{-2}&=&\sum_{s=-2(2^L-1)}^{2^{L+1}-1}Q(s)
                                         (2s-1)^{-2}\cr
&=&\sum_{s=1}^{2^{L+1}-1}\frac{Q(s)+Q(1-s)}{(2s-1)^2}\;.
\end{eqnarray}
Using a simple counting argument based on the register principle
\cite{Mattson}, one can show that
\begin{equation}                                                \label{baker42}
Q(s)+Q(1-s)=2^{L+1}-s+\frac{1}{2}[1-(-1)^s]  \;,
\end{equation}
from which one obtains
\begin{eqnarray}                                                \label{baker43}
\sum_{s=1}^{2^{L+1}-1}\frac{Q(s)+Q(1-s)}{(2s-1)^2}&=&
\sum_{s=1}^{2^L-1}\frac{2^{L+1}-s}{(2s-1)^2}-\sum_{t=1}^{2^L-1}(4t-1)^{-2}\cr
&=&2^L[\,\frac{\pi^2}{4}-O(\frac{L}{2^L})]\;,
\end{eqnarray}
where we have used the relations
\begin{equation}                                                \label{baker44}
\sum_{s=1}^{2^{L}}(2s-1)^{-2}=\frac{\pi^2}{8}+O(2^{-L})\;,
\hspace{1cm}
2^{-L}\sum_{s=1}^{2^{L}}\frac{1}{2s-1}=O(\frac{L}{2^{L}})\;.
\end{equation}
Combining Eqs.~(\ref{baker40}), (\ref{baker41}) and (\ref{baker43}), we obtain
Eq.~(\ref{baker39}) as required. We now  rewrite Eq.~(\ref{baker39}) 
in the symmetric form
\begin{equation}                                                \label{baker45}
\Tr[\rho_k\hat{B}\rho_{k-1}\hat{B}^{\dag}]=
2^{-r}[1-O(\frac{r}{2^{r-k}})]
\end{equation}
and introduce the distance measure between density matrices induced by 
the Eucledian norm \cite{HornJohnson},
\begin{equation}                                                \label{baker46}
d(\rho,\rho')\stackrel{\rm def}{=}\sqrt{\Tr(\rho-\rho')^2}\;.
\end{equation}
This distance measure is unitarily invariant and obeys the triangle
inequality. We will now prove that (\ref{baker45}) and (\ref{baker46}) 
imply 
\begin{equation}                                                \label{baker47}
d(\rho_k,\hat{B}^k\rho_0[\hat{B}^\dag]^k)=
                                  O(2^{\frac{k}{2}-r}\sqrt{r})\;.
\end{equation}
Using the cyclic property of the trace, we have
\begin{equation}                                                \label{baker48}
d(\rho_k,\hat{B}\rho_{k-1}\hat{B}^{\dag})=\sqrt{\Tr\rho_k^2+
\Tr\rho_{k-1}^2-2\Tr(\rho_k\hat{B}\rho_{k-1}\hat{B}^{\dag})}\;.
\end{equation}
Since $\Tr\rho_k^2=2^r/2^{2r}=2^{-r}$ for any $k$,
\begin{equation}                                                \label{baker49}
[d(\rho_k,\hat{B}\rho_{k-1}\hat{B}^\dag)]^2=
 2^{-r+1}-2\Tr(\rho_k\hat{B}\rho_{k-1}\hat{B}^\dag)\;,
\end{equation}
which, together with Eq.~(\ref{baker45}), implies
\begin{equation}                                                \label{baker50}
d(\rho_k,\hat{B}\rho_{k-1}\hat{B}^\dag)=
            O(2^{\frac{k}{2}-r}\sqrt{r})\;.
\end{equation}
The case $k=1$ of (\ref{baker47}) follows directly from (\ref{baker50}).
Assuming that (\ref{baker47}) is true for a given value of $k$
and using the unitary invariance of the distance (\ref{baker46}), we have
\begin{equation}                                                \label{baker51}
d(\hat{B}\rho_k\hat{B}^\dag,\hat{B}^{k+1}\rho_0
         [\hat{B}^\dag]^{k+1})= O(2^{\frac{k}{2}-r}\sqrt{r})\;.
\end{equation}
Substituting $k+1$ for $k$ in Eq.~(\ref{baker50}), we get
\begin{equation}                                                \label{baker52}
d(\rho_{k+1},\hat{B}\rho_{k}\hat{B}^\dag)=
O(2^{\frac{1}{2}(k+1)-r}\sqrt{r})\;.
\end{equation}
Using the triangle inequality for the distance measure (\ref{baker46}),
it follows from (\ref{baker51}) and (\ref{baker52}) that 
\begin{eqnarray}                                                \label{baker53}
d(\rho_{k+1},\hat{B}^{k+1}\rho_0[\hat{B}^\dag]^{k+1})&=&
                O(2^{\frac{k}{2}-r}\sqrt{r})+
                O(2^{\frac{1}{2}(k+1)-r}\sqrt{r})\cr
&=&O(2^{\frac{1}{2}(k+1)-r}\sqrt{r})\;.
\end{eqnarray} 
By induction, this completes the proof of (\ref{baker47}) for any
$k\leq k_{\rm max}$.
On the other hand
\begin{equation}                                                \label{baker54}
d(\rho_k,\hat{B}^k\rho_0[\hat{B}^\dag]^k)=
      \sqrt{\Tr\rho_k^2+\Tr\rho_0^2-2\Tr(\rho_k\hat{B}^k
                                    \rho_0[\hat{B}^\dag]^k)}\;,
\end{equation}
hence using Eq.~(\ref{baker47}) it follows that 
\begin{equation}                                                \label{baker55}
\sqrt{2^{1-r}-2\Tr(\rho_k\hat{B}^k\rho_0[\hat{B}^\dag]^k)}=
O(2^{\frac{k}{2}-r}\sqrt{r})\;,
\end{equation}
and finally
\begin{equation}                                                \label{baker56}
\Tr(\rho_k\hat{B}^k\rho_0[\hat{B}^\dag]^k)=
2^{-r}[1-O(\frac{r}{2^{r-k}})]\;,
\end{equation}
which is equivalent to (\ref{baker38}) as required.

\subsection{Atypical initial states}

In this section, we show that the state
$|0^r\y^1.\y^2\ket$, where $0^r$ is a string of $r$ zeros, is an atypical state
in the sense of the discussion at the end of Sec.~\ref{sec:results}, i.e., 
we show that the state $|0^r\y^1.\y^2\ket$ satisfies Eq.~(\ref{baker21}).
A direct calculation gives
\begin{eqnarray}                                                \label{baker57}
\Tr[P^{r,1}_{\ys}\hat{B}|0^r\y^1.\y^2\ket\bra 0^r\y^1.\y^2|
               \hat{B}^{\dag}]
&=&\sum_{|\xs|=r-1}\sum_{g=0}^{1}|C^{\rm 1st}(0^r\y^1\y^2,\x\y^1\y^2g)|^2\cr
&=&\frac{8}{\pi^2}\sum_{v=0}^{2^{r-1}-1}\frac{1+O(4^{r-n})}{(4v+1)^2}\;.
\end{eqnarray}
Substituting $t=2v$, we have
\begin{eqnarray}                                                \label{baker58}
\Tr[P^{r,1}_{\ys}\hat{B}|0^r\y^1.\y^2\ket\bra 0^r\y^1.\y^2|
\hat{B}^{\dag}]&=&\frac{8+O(4^{r-n})}{\pi^2}\sum_{t=0}^{2^r}
                   \frac{1+(-1)^t}{2(2t+1)^2}\cr
               &=&\frac{4+O(4^{r-n})}{\pi^2}
                  \left(\sum_{s=1}^{2^r+1}(2s-1)^{-2}+
                   \sum_{t=0}^{2^r}\frac{(-1)^t}{(2t+1)^2} \right)\;.
\end{eqnarray}
Using Eq.~(\ref{baker44}) and the series representation
of Catalan's constant $G\simeq0.915965$ \cite{GradshteynRyzhik},
\begin{equation}                                                \label{baker59}
G=\sum_{t=0}^{\infty}\frac{(-1)^t}{(2t+1)^2} \;,
\end{equation}
it follows that
\begin{eqnarray}                                                \label{baker60}
\Tr[P^{r,1}_{\ys}\hat{B}|0^r\y^1.\y^2\ket\bra 0^r\y^1.\y^2|
\hat{B}^{\dag}]&=&\frac{\pi^2+8G}{2\pi^2}+O(4^{r-n})+O(2^{-r})\cr
               &\simeq&0.871+O(4^{r-n})+O(2^{-r})\;.
\end{eqnarray}
Since one can treat $n-r$ and $r$ as independent variables,
this expression can be made smaller than $0.872$ by choosing $n-r$ and $r$ large
enough. For the initial state $|0^r\y^1.\y^2\ket$, the asymptotic
relation (\ref{baker18}) is thus violated.

\section{Acknowledgements}

Thanks to C. M. Caves, C. A. Fuchs and M. Saraceno for useful
discussions. We acknowledge the hospitality of the Isaac Newton
Institute in Cambridge, where part of this work was completed. 
This work was supported in part by the 
US Office of Naval Research (Grant No.~N00014-93-0-0116) and by the UK 
Engineering and Physical Sciences Research Council.

\end{document}